# A "Perspectival" Mirror of the Elephant: Investigating Language Bias on Google, ChatGPT, YouTube, and Wikipedia

Queenie Luo
Harvard University
queenieluo@g.harvard.edu

Michael J. Puett
Harvard University
puett@fas.harvard.edu

Michael D. Smith
Harvard University
mike_smith@harvard.edu

## ABSTRACT

Contrary to Google Search's mission of delivering information from "many angles so you can form your own understanding of the world," we find that Google and its most prominent returned results – Wikipedia and YouTube – simply reflect a narrow set of culturally dominant views tied to the search language for complex topics like "Buddhism," "Liberalism," "colonization," "Iran" and "America." Simply stated, they present, to varying degrees, distinct information across the same search in different languages, a phenomenon we call language bias. This paper presents evidence and analysis of language bias and discusses its larger social implications. We find that our online searches and emerging tools like ChatGPT turn us into the proverbial blind person touching a small portion of an elephant, ignorant of the existence of other cultural perspectives. Language bias sets a strong yet invisible cultural barrier online, where each language group thinks they can see other groups through searches, but in fact, what they see is their own reflection.

## KEYWORDS

Language Bias, Google, ChatGPT, YouTube, Wikipedia

## 1 INTRODUCTION

What is your impression of Buddhism? To Westerners, Buddhism is generally associated with spirituality, meditation, and philosophy, but people who primarily come from a Vietnamese background might see Buddhism as closely tied to the lunar calendar, holidays, mother god worship, and capable of bringing good luck. One from a Thai culture might regard Buddhism as a canopy against demons, while a Nepali might see Buddhism as a protector to destroy bad karma and defilements. In the hope of finding a thorough *global* picture of Buddhism, we type in "Buddhism" on Google. However, instead of helping us to solve the paradox, Google's top 50 search results simply reflect these distinct parochial cultural impressions about Buddhism as we switch the language for our query, where an English search returns a list of meditation centers and education channels, but a Vietnamese search returns large Vietnamese Buddhist organizations' sites listing Buddhist ceremonies like making offerings and praying for peace.

It seems *language* is a distinctive feature in filtering information on Google. Living in a highly globalized era and with the advent of the internet, it is easy to assume that the information displayed by our search engines ranks all websites across the globe. Indeed, Google's mission reaffirms this assumption: "For more complex topics, Search is a tool to explore many angles so you can form your own understanding of the world."[1] However, contrary to Google's mission of presenting *many angles* of the world, its results reflect a narrow set of cultural attitudes and values tied to the search language. We use language as an *instrument to communicate* with Google to acquire information, but Google often treats language as a *navigator to orient* our cultural position.

The Hindu fable of the blind men and the elephant poignantly points out the pitfall of limited perception. Six blind men approach an elephant. One person touches its tail and says an elephant is like a rope. The second person touches its trunk and says it is like a snake. A third person touches its leg and says it is like a tree. Another touches its side and says it is like a wall; another its ear and says it is like a fan; and another its tusk and says it is like a spear. Even though the blind men are being truthful with their experiences, their understanding of an elephant is incomplete because of their narrow perspective. Their understanding of the elephant is what Nietzsche would call *perspectival*.

We often derive our experience and knowledge about the world from the tiny frame of a search window. We type a word or question in a search box – Google in our case – to obtain some sort of wisdom about a thing or person, and we tend to believe the search results are the totality of the concepts and views. Sadly, rather than showing us a global picture of the 'elephant,' Google directs us to different parts of the 'elephant' based on the language we use. The language we use ends up as a cultural filter to perpetuate ethnocentric views, where a person evaluates other people or ideas based on their own culture.

This phenomenon is not unique to Search as we find that YouTube and Wikipedia, which are featured prominently in Google's search results, also exhibit discrepancies in the information they present when we vary the language used to access their corpus.

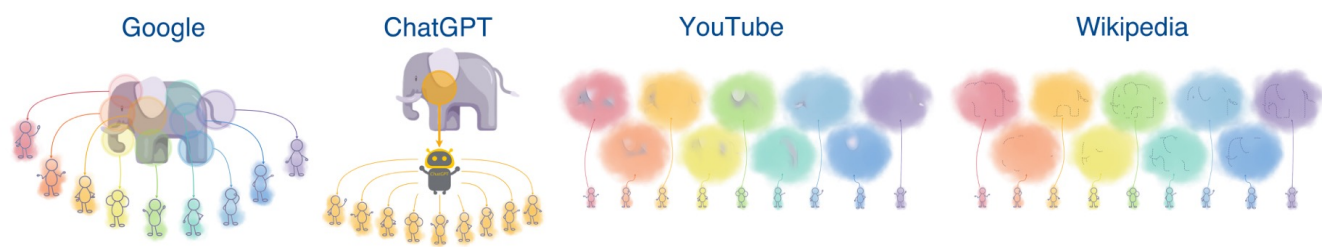

**Figure 1 This figure illustrates how various language communities access information via Google, ChatGPT, YouTube and Wikipedia. The elephant symbolizes the global knowledge of a complex concept available online. Each individual represents a distinct language community, while the colors represent information they receive through these platforms. Each language community receives their culturally dominant views on Google. As ChatGPT is predominantly trained on English data, the Anglo-American perspective overtly dominates its outputs. YouTube videos carry images and sounds that construct a deeper ethnocentric experience for users than Google and Wikipedia's textual information. In that respect, YouTube videos zoom in to the 'color' and 'texture' of the elephant, but the 'color' and 'texture' vary by the user's search language. Information on Wikipedia is confined to a scholarly perspective, where belief-oriented materials are contextualized as cultural phenomena. In that respect, Wikipedia presents a rough sketch of the elephant with its 'color' and 'texture' missing.**

However, things turned out to be different with ChatGPT (Feb 13, 2023 Version). As it is primarily trained on English language data,[2],[3] it presents the Anglo-American perspective as the normative view, reducing the complexity of a multifaceted issue to the single Anglo-American standard; non-English perspectives are unknowingly dismissed as unimportant, irrelevant, or wrong without critical examination. (Figure. 1)

In this paper, we investigate *language bias* on Google, YouTube, and Wikipedia – three popular platforms commonly used to expand our understanding of the world around us – to examine whether language yields similar or different information on a complex topic.[1] Since ChatGPT has been integrated into Bing and is rapidly growing in popularity. We provide a brief analysis of ChatGPT as an extension to Google Search.

We say that a platform exhibits language bias if the *language* of a query leads to a systematic deviation in sampling information that prevents it from accurately representing the true coverage of topics and views available in their repository. The term "bias" has diverse meanings and interpretations across academic fields. Our use of bias in this paper is analogous to *statistical sampling bias*. For example, on the English search term "wedding dress," Google Image primarily shows the white gown type of Western-style wedding dress rather than a diverse global spectrum of culturally-specific bridal dresses, such as a Japanese kimono-style bridal dress or an Indian sari-like draped bridal garment. We say that its ranking algorithm exhibits bias towards websites written in the query language, resulting in a significantly higher probability of being presented than non-query-language websites on Google. Our goal is not to quantify the statistical discrepancy between the results and the entire repository (as it can be quite difficult), but to identify whether a discrepancy exists and begin to understand its causes.

We begin our work with a focus on the complex topic of Buddhism. Buddhism has deep roots in different language communities around the world, and some of our authors have the academic training necessary to assess whether we have seen a relatively comprehensive overview of Buddhism based on our search queries. Our study also shows that language bias is not unique to Buddhism but applies to a wide range of terms where culture matters, such as political ideologies like "Marxism" and "Liberalism," historic events or political movements like "Colonization" and "Cultural Revolution," and geographical names like "Iran" and "Russia" and "America." Our analyses of these terms support our argument, but in addressing the social implications of these findings, this paper focuses on the contentious area of political ideologies. We also provide a brief discussion on terms that have not been strongly influenced by language bias, such as "Automatic Differentiation," "Jacobian Matrix," and "Hidden Markov Chain." Topic coverage and views on these terms do not deviate much across different language searches.

If we see a search engine (Figure. 2), Google in our case, as a world container of billions of indexed websites, Wikipedia and YouTube are two prominent pieces of the constituents that contribute to the entirety of knowledge online. The new search tool, ChatGPT in our study, can be seen as an intelligent assistant or delegator of a search engine. Language bias exists in almost all spheres, but it is further filtered by platform orientations, resulting in varying manifestations and impacts on users.

---

[1] Google's mission statement uses this term – "complex topic."

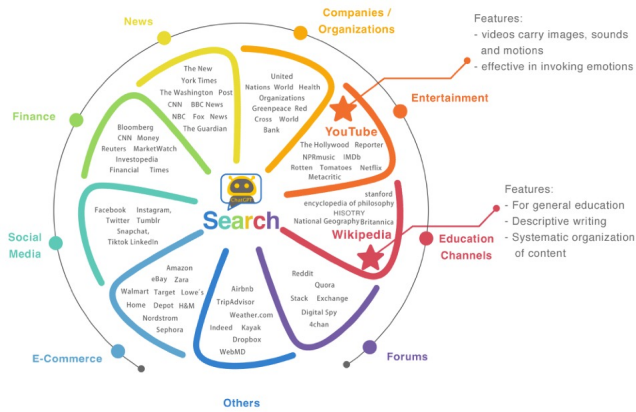

**Figure 2 This figure illustrates the internet sphere. Considering a search engine as indexing billions of websites, YouTube and Wikipedia are two key contributors to online knowledge. ChatGPT, a new search tool, serves as an intelligent search engine assistant. Language bias is prevalent in nearly every domain, but it is further filtered by platform orientations, resulting in varying manifestations and impacts on users.**

We find that Google's top search results show a strong preference towards the dominant cultural phenomena of a language community, which can be influenced by many factors such as the majority's interests, powerful institutions, and major social events. As a new search tool, ChatGPT helps its users quickly find and summarize relevant information. It acts *like* a delegatory sighted person describing what an elephant looks like to its clients and covered itself as a multilingual chatbot, but in fact, it is almost 'blind' to non-English sources and primarily 'thinks' in English. We find that the ChatGPT hosted on OpenAI.com (February 13, 2023, version) presents the Anglo-American perspective as truth, largely ignorant of the existence of other cultural perspectives. The one being integrated into Bing (Accessed March 1, 2023) responds in the input languages with the same sentence structure but conveys different perspectives and messages.

Switching to the popular sites featured on Google, language bias on YouTube is similar to Google's but amplified. YouTube's ranking system seems largely determined by the video's freshness and popularity. Popularity seems to be related to users' interests and satisfaction, with videos that align with more people's interests or preexisting cultural beliefs receiving more likes and views and being promoted through the recommendation section on YouTube. In contrast to text-based information, YouTube videos carry images, sounds, and motions, making them more effective in evoking emotional resonance. A music performance of a blind Tibetan girl chanting a Buddhist mantra[4] transports the suffering and sorrow in her life as well as the courage and strength given by her faith. The smile and serenity expressed on a Buddhist practitioner's face[5] show the power of Buddhist teachings. This dynamic use of audio and visual information creates a deeper connection between viewers and videos than textual information, constructing a deeper ethnocentric experience for viewers. In that respect, YouTube videos zoom in to the 'color' and 'texture' of the elephant, but the 'color' and 'texture' vary by the user's search language (Figure 1).

In contrast, information on Wikipedia is confined to a scholarly perspective that belief-oriented materials such as songs, personal testimonies, dreams, and diaries are contextualized as religious or cultural phenomena, rather than aimed at persuading its audiences or invoking certain emotions. In that respect, Wikipedia presents a rough sketch of the elephant with its 'color' and 'texture' of missing, where the color and texture stand for emotions and personal experiences (Figure. 1). As suggested in Wikipedia's five pillars,[6] Wikipedia is an encyclopedia that provides summaries of knowledge and is written from a neutral point of view. We find that even though the tones of voice and views do not differ much in Wikipedia articles across languages, *topic coverage* in Wikipedia articles tends to be directed by the dominant intellectual traditions and camps across different language communities, e.g., a French Wikipedia article focuses on French thinkers, and a Chinese article stresses on Chinese intellectual movements. Wikipedia's fundamental principles or objectives filter language bias, making it heavily rely on intellectual and academic institutions.

The intangible biases in shaping the perception of the users should not be overlooked or understated. In some versions of the fable, the blind men suspect each other of being dishonest and fall into violent conflicts. While in other versions of the fable, an *authentic* sighted man sees the entire elephant and explains to the six blind men that their experience is true but incomplete because each of them only touches a portion of the elephant. The six blind men reconcile with joy. The appearance of a *real* sighted man solves the paradoxes and conflicts between the blind man.

We somehow treat a search tool or AI bot as a sighted person, imputing authoritativeness to it to show us the world beyond our physical limits. However, can a search tool or AI bot become a *reliable* sighted person, helping us piece together a genuine global image of the elephant and solve paradoxes? Should they aspire to do that? What roles should they take on in our highly globalized era?

A tech company's profitability does not naturally align with higher social and moral obligations. Aligning search results with the majority's views of their own culture will increase user's immediate satisfaction and maximize profit through more ad clicks and purchases. But this in fact becomes an insidious engine of 'tyranny of opinion' and 'tyranny of the majority' characterized by uniformity and suppression of unpopular and counter opinions. Highlighting voices of privileged groups such as powerful corporations or agencies might make Google look authoritative and trustworthy, but could in turn undermine the true values of democracy, free speech, and equity. Voices from grassroots institutions can hardly be heard through Google.

In fact, Google's algorithm constructs hyper-focused but varied realities where citizens think the space is singular and global but there are deliberate human choices built into the system that manipulates what we *can* see and *should* see online. Google, as it turns out, becomes a handy tool for powerful corporations and private hands to constitute "objective information" online that is, de facto, skewed in their favor. What Google promises us to do – presenting many angles of the world – is different from what it actually does – reinforcing dominant local views in a language community. Its algorithm perpetuates dominant *local* linguistic views in a *global* information space, constructing a seemingly coherent, agreeable, and interconnected world. The profit-driven technology has served as a facilitator of social alienation rather than cohesion.

The issue of how a search engine should rank websites and display information has been largely assumed as technical and business problems. However, the question of what should be ranked highly or deemed as irrelevant or wrong, especially on complex topics like "Buddhism" and "liberalism," is a deeply complex issue. Similar to the emerging AI bots, what's the best way to answer the question "what is liberalism?" has no easy solution. This demands a collective effort involving experts from diverse fields.

People have paid much attention to the potential existential threat posed by AI. However, what's more pertinent is the insidious force of algorithms and AI models exerting through information distribution, content rating, and website rankings.

## 2 DATA AND EVALUATION METRICS

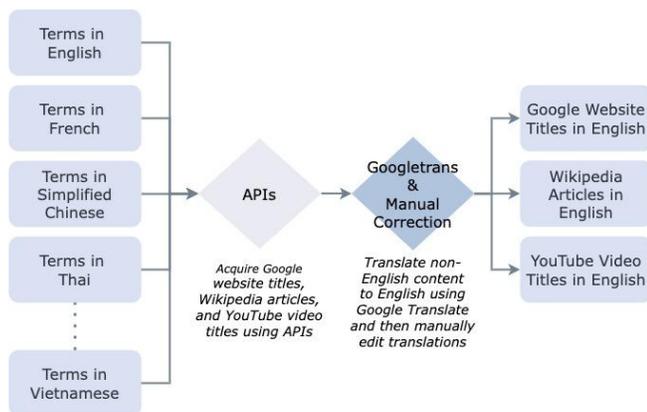

**Figure 3 This figure shows the data processing pipeline. We first selected 13 common English terms in Buddhism and translated them into 12 languages. Then, we used APIs to acquire data on Google, YouTube, and Wikipedia, and we used Google Translation to translate data into English and manually edited them. ChatGPT's data acquisition process is excluded from this pipeline because we used questions rather than search terms.**

## 2.1 Selecting and Translating Search Terms

To investigate discrepancies on Google, ChatGPT, YouTube, and Wikipedia, we selected some common English terms in Buddhism, such as "Buddhism," "Buddhist meditation," and some well-known Buddhist leaders like the "Dalai Lama." These terms tend to appear under "People also ask" and "Related Searches." The full table can be found in Appendix 4.

Next, we translated the selected English terms into 12 languages across Asia and Europe.[2] Buddhism has had cultural significance in many Asian countries in history, such as China, Thailand and Vietnam, and currently has millions of followers there. Comparatively, Buddhism only started to attract the attention of Western intellectuals in the latter half of the 19th century, and gained popularity during the 20th century. Comparisons between Asian and European language contents allowed us to assess the types of biases that existed with respect to their cultural and political contexts.

In addition, to study language bias beyond Buddhism-related terms, we selected a wide range of subjects, including political ideologies like "Liberalism" and "Marxism," major historical events or political movements like the "Cultural Revolution," "Colonization" and the "Feminist Movement," as well as geographical names like "Russia," "Iran" and "America." We also translated these terms into 12 languages to acquire data on these platforms.

## 2.2 Acquiring Data

To acquire Google search results, we used the SerpApi[7] to obtain our Google search results, which allowed users to specify maximum page numbers and geographical locations to scrape results from the Google Local Services search page. According to Chitika, "sites listed on the first Google search results page generate 92% of all traffic from an average search," and only 1.2% of users ever click on the third page.[8] Recently, Google has changed to a single continuous page layout so that users no longer have to click through multiple pages and are able to view more content through scrolling. Thus, we collected the top 50 website titles on Google. Since IP address might still be used as a ranking factor in Google's ranking algorithm,[9] we set the location as "U.S." We chose *not* to switch location to a language's home country, say setting the location as "Japan" while using Japanese terms, because our U.S. IP address and other potential factors will still influence the search results regardless of our location setup, so we kept our location setting as "U.S." as an attempt to study users speaking different languages in the U.S. Similarly, we used SerpApi to acquire top 35 YouTube video titles across languages with location set to "U.S." The top 35 videos include videos under "People also watched" and "From related searches."

We used Wikipedia-API[10], a python wrapper for

---

[2] We use Simplified Chinese for all Chinese-related searches in this paper.

Wikipedia's API, to get Wikipedia pages. We acquired Wikipedia articles from language-specific Wikipedia sites by setting the corresponding language codes in the API. For example, Japanese Wikipedia articles were acquired from ja.wikipedia.org, and French articles from fr.wikipedia.org.

During the time this research was conducted (March 2022 - February 2023), there were two versions of ChatGPT and an API was not yet available. The first one was the one hosted on OpenAI.com launched in November 2022. We refer to this model (Feb 13 version) as ChatGPT-OpenAI in this paper. And the second one was the one being integrated into Bing search in February 2023, and we call this model as ChatGPT-Bing. Data for both ChatGPT models was collected manually and compared against our findings for Google Search.

## 2.3 Translating Data

In this step, we translated all acquired data from Google, ChatGPT, Wikipedia, and YouTube into English using Googletrans[11], a python library that implements Google Translate API. We translated all data into English for comparison. Due to the presence of technical terminologies in Buddhist studies and other complex topics, translations are further edited manually.

## 2.4 Evaluation metrics

We measured information discrepancies using the Term Frequency-Inverse Document Frequency (TF-IDF) + cosine similarity. TF-IDF is a statistical metric that computes word vector representations based on word frequency in a document with respect to other documents in a corpus. It has been widely used in topic classification and has proven to produce robust results. However, there is no perfect metric to measure "similarity" because we do not have a mathematical definition of "similarity." The advantage of using the TF-IDF algorithm over other large pre-trained NLP models, such as BERT or Word2Vec, is that they are relatively simple and easy to interpret and avoid an additional layer of errors and bias that come from pre-trained models.

In this study, we compute the TF-IDF similarity score using Google website titles and YouTube video titles. On search term $T$, a low similarity score between language $X$ and $Y$'s search results indicates language communities of $X$ and $Y$ see highly different information on search term $T$. It is important to note that we do *not* take the numerical value of the similarity score *literally* because TF-IDF is sensitive to word frequency that a word with uncommonly high frequency will produce overall higher similarity scores with respect to other language contents. For example, if French and English search results have a similarity score of 0.68 and German and English search results have 0.66, we say that both French and German have about the same level of similarity to English search results, and we do not say that French is more similar to English than German.

Unlike Google website titles and YouTube video titles that have similar lengths across languages, Wikipedia articles vary in length and ChatGPT can generate multiple slightly different passages for the same question. For these reasons, we do not compute TF-IDF similarity scores for Wikipedia articles and ChatGPT outputs, but instead provide specific examples and detailed analytical discussions for them.

## 3 RESULTS & INTERPRETATIONS

In this section, we present an in-depth analysis of language bias on Google, YouTube, and Wikipedia on Buddhism-related terms. We also provide a brief discussion of ChatGPT at the end of the Google section, as it is touted as the next generation of Search.

## 3.1 Buddhism-related search results on Google

Over the course of 2,500 years, Buddhism which originated in India has taken roots in virtually every country in Asia and has spread to Western countries in recent times. A diverse Buddhist tradition flourished globally, and we expect that everyone who uses the internet can see such a *diverse tradition* of Buddhism on the web. Thus, we start our analyses with Google Search because it is generally the starting place for most people to surf the web (as of February 2023).

### 3.1.1 On the term "Buddhism"

Using the search term "Buddhism," we acquired Google's top 50 website titles across 12 different languages. Our TF-IDF cosine similarity metric shows that there are differences across *all* language pairs in Figure. 4. That is, no two languages' search results are distinctively similar to each other. However, we do see a cluster between European languages, which happens around the threshold of 0.6 in this chart.

Looking into the websites, the common topics on "Buddhism" across all languages' searches results included "Buddhism basic beliefs" and "Buddhism history," but besides these common topics, European languages search results had more websites on philosophy, psychology, therapy, meditation, and life wisdom, and asking whether Buddhism is a philosophy or religion. Several English search results were related to retreat centers and foundations such as tricycle.org, lamayeshe.org, and education channels like history.com and scientificamerican.com. The French and German search results also focused on educational sites, such as websites that were encyclopedic in nature or deal with world history, and various Buddhist institutions.

Asian language search results presented different realities about Buddhism. Several Chinese websites were hosted under official government websites, stating the Party's policies over how Buddhist monasteries should operate in China – 'monastic education should be strengthened,'[12] and the Party's official attitudes towards Buddhism – 'Buddhism as a source of Chinese cultural confidence,'[13] 'a core value in Mahayana Buddhism is altruism and helping people in need.'[14] Vietnamese also had government-generated pages that address Buddhism development in Vietnam, websites that contain life stories as a Buddhist mendicant, and various Buddhist organizations showcasing ritual activities and meditation practices in Vietnam. As other minority languages lack trustworthy websites, the results rank Facebook posts highly. Thai search results contained a Facebook page explaining the difference between Buddhism and ghost religion, and the Nepali contents consisted of a large number of pdfs rather than actual websites.

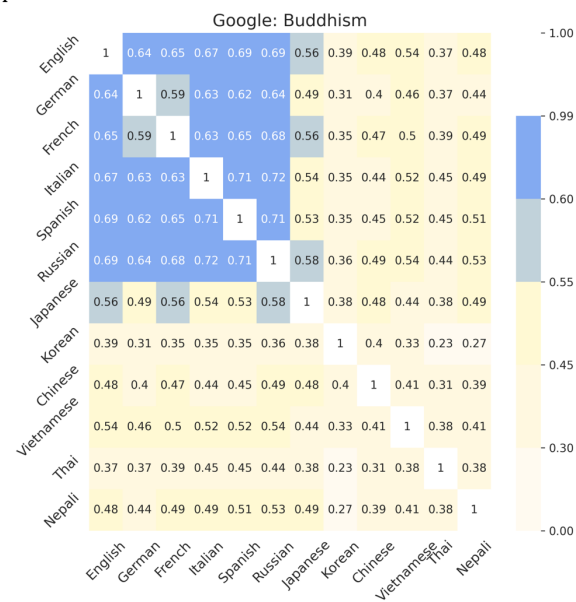

**Figure 4** This figure shows the TF-IDF cosine similarity scores for Google's top 50 website titles on "Buddhism." It presents a cluster between European languages, which happens around the threshold of 0.6.

These different pictures presented on Google are consistent with how Buddhism is being portrayed in Western and Asian communities. Buddhism in the West, or the so-called "Western Buddhism," refers to a form of Buddhism transmitted in North American and European countries that has been heavily commodified, scientized and institutionalized. [15],[16] Nowadays in the West, a Buddha statue is more often used as a decoration, rather than a relic to be worshiped. "Mindfulness meditation" is taught in hospitals and studied in neurology laboratories; it has been branded as spirituality and self-healing in a secular form.[17],[18] And numerous large foundations and initiatives offer Buddhist teachings for calm and insight rather than religious practices.[19]

On the other side of the globe, China's Communist Party follows the traditional Marxist view on religion that religion is counterproductive to social change. However, rather than removing religion through force, the Party allows religion to stay but based on the government's approval.[20] The presence of the government-generated websites, articles and news about Buddhism shows the Party's rigorous regulations over Buddhism in China. Vietnamese Buddhism is well known for its grassroots Buddhist organizations of advocating nonviolence, mindfulness practice and engagement in social actions. Thich Nhat Hanh, a prominent Vietnamese Zen Buddhist monk, peace activist, and author, is a key figure in this movement. He has played a significant role in promoting the concept of "Engaged Buddhism," which seeks to apply Buddhist teachings to social, political, and environmental issues.[21], [22] The rise of large Buddhist organizations, such as the Unified Buddhist Church (UBC), was a response to protest against the Vietnam war. A large portion of websites returned using Vietnamese terms end with '.org' or '.net,' indicating that they are backed by Buddhist organizations. Their contents endorse a positive image of Buddhism and attempt to correct 'misconceptions' of Buddhism – 'Buddhism is as practical as science,' and 'Buddhism is considered as a philosophy.'[23], [24]In the intellectual milieu of Buddhism, whether Buddhism is 'scientific' or considered as philosophy are hotly debated questions,[3] but views in the Vietnamese websites are largely dominated by one side of the argument. In Thai culture, spirits, ghosts, and otherworldly beings are important elements. Thai people have the tradition of building miniature spirit houses outside homes or on the street to appease or build relationships with spirits,[25],[26] so it is not surprising to see elements of spirits or ghosts appear in Thai searches. Lastly, Buddhism in Nepal shares a close religious tradition with Indo-Tibetan Buddhism, and expectedly the search results using Nepali terms return contents related to Indian Buddhism and Tibetan cultural symbols.

Rather than presenting a pluralistic global picture of Buddhism, Google's top search results show a strong preference towards the dominant cultural phenomena of Buddhism in a language community. We use the word "strong" here because some dominant topics in one language's top search results almost *never* appear in other languages. Suppose the major topics on Buddhism contain {A, B, C, D, E, F, G}, English's top search results only get {A, B, C, D}, Chinese {A,

---

C, E, G} and Vietnamese {A, B, E, F}. English users never see {E, F, G} because this information is not written in English or discussed in English websites (see Figure. 5).

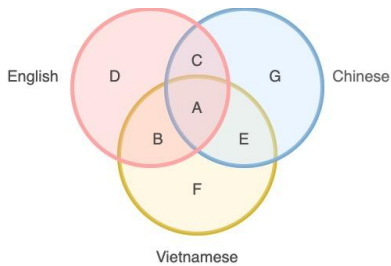

**Figure 5** This figure shows a Venn diagram of dominant topics (A-G) in Google's top search results for English, Chinese, and Vietnamese queries. This diagram illustrates the unique and overlapping topics in each language's top search results, highlighting information discrepancies between languages. Despite the fact that all language search results cover topic A, their perspectives and opinions expressed in this topic are different and tend to align with their mainstream cultural views.

In addition to differences in topic coverage, even when websites in English, Chinese, and Vietnamese cover the same topic (Topic A), the perspectives and opinions expressed in the content tend to differ across languages. This is primarily because the views and discussions tend to align with the mainstream cultural values within each respective language community. These cultural differences can result in varying narratives and interpretations of the same topic.

We find that the dominant cultural phenomena can relate to many factors. Based on the data we demonstrated above, in some cases, the dominant cultural phenomena could relate to the majority's interests, like European language users are more likely to be interested in Buddhist philosophy and meditation rather than ritual practices, resulting in sites of philosophy and meditation being ranked highly in European languages' searches. In other cases, the dominant cultural phenomena could relate to policies or objectives of powerful agencies or institutions that are skilled in Search Engine Optimization (SEO) to make their web pages rank highly. For example, the top returned Chinese government sites and Vietnamese organizations' sites aim to inculcate and inform the majority. Following the majority or maintaining the status quo might not be their primary goal. Occasionally, the dominant cultural phenomena could be some major socio-political or religious events, such as the Buddha's birthday.

### 3.1.2 Popular narrower topics related to Buddhism on Google
We then narrowed down our search from "Buddhism" to some popular narrower concepts, such as "Buddhism belief" and "Buddhism practices" because these terms tend to appear under "People also ask" and "Related searches." We find that language bias exists in all Buddhism-related terms but is especially prominent for belief-oriented concepts, such as "karma in Buddhism" and "Gautama Buddha."

The law of karma says: for every action we make, there is an equal and opposite reaction we receive. Differing from many monotheists' belief in God's rewards and punishments, karma tells us we are the creators of our own good or bad fortune. It is a fundamental Buddhist worldview closely tied to Buddhist ethics and moral conduct. "Gautama Buddha" refers to the founder of Buddhism, Siddhartha Gautama. Rather than having a single creator God, there are many Buddhas in the Buddhist pantheon. For both terms, our TF-IDF cosine similarity metric shows an apparent cluster for European languages but no prominent cluster for Asian languages. The cluster threshold happens around 0.7 for "Karma in Buddhism" and 0.8 for "Gautama Buddha."

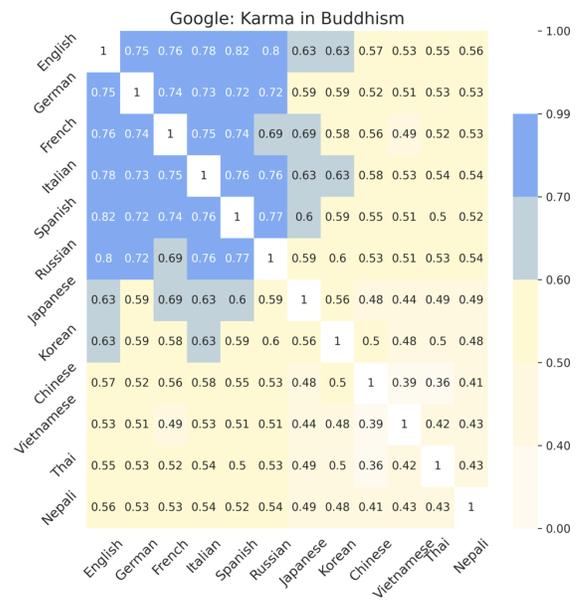

**Figure 6** This figure shows the TF-IDF cosine similarity scores for Google's top 50 website titles on "Karma in Buddhism." It presents a cluster between European languages, which happens around the threshold of 0.7. Japanese website titles tend to be closer to European language contents than other Asian language contents.

Looking into the website titles, European language search results were centered around the definitions and theories of karma, i.e., 'any kind of intentional action is regarded as karma.' Comparatively, besides talking about the basic definitions of karma, Asian language search results also touched on topics like 'how to resolve bad karma and accumulate good karma' and stories of karmic retribution and reward. Using Vietnamese search on Google, eight out of ten returned websites on the first-page talk about 'how to eliminate/resolve/neutralize karma.' Chinese search results had an article placed on around the 10th on Google that explains why a pious kindhearted grandma got into a car accident based on the law of karma.[27] A number of Thai search results mentioned Phra Vessantara[28],[29] which is the story of Prince Vessantara, the last incarnation of Gautama Buddha. The story is generally told during an important Thai festival Khao Phansa to explain how karma works.

It seems Google's top search results get even more ethnocentric for narrower topics, and its ranking algorithm incorporates a set of *assumptions* based on search language. Since Buddhism is a foreign religion to most westerners, most European language users are unfamiliar with Buddhist concepts. Websites that give descriptive or theoretical definitions of 'karma' are ranked highly and websites that discuss the 'workings of karma' are demoted. For example, the English websites "18 Ways to Create Good Karma"[30] and "Buddhist Laws For Karmic Cleansing – Exploring your mind"[31] that teach you how to work with karma are found on the 150th on Google. But similarly focused websites in Asian languages are ranked highly.

Searching for "Gautama Buddha," European language results focused on 'who is the Buddha' and 'the life of the Buddha,' while Chinese results asked about 'the difference between Amitabha Buddha and Gautama Buddha' and 'the Bodhi Day.' The Amitabha Buddha is a popular icon in current Chinese society because it frequently appears in Chinese art, literature, and entertainment. The Bodhi day is an important Chinese Buddhist holiday that commemorates the historical Buddha's attainment of enlightenment. Both Vietnamese and Thai search results touched on the tales of the Gautama Buddha, including his legendary birth[32], because the humanness of the historical Buddha is emphasized in Theravada Buddhism.[4]

For belief-oriented terms, or terms that involve a deeper engagement with Buddhism, the gap between European and Asian language content increases. European language contents tend to ask 'what it is,' inquiring with a complete mental detachment from the concept itself and inviting the readers to contemplate, but Asian language search results ask 'how to solve the problem in my life,' imposing a settled opinion of correctness and emphasizing on its worldly implications and moral consequences.

Searches on public figures also bore similar patterns across different language contents. Chögyam Trungpa is a Buddhist scholar and the founder of many meditation centers in Europe and the U.S. He attracted a large number of western students to Tibetan Buddhism during the 1970s and 1980s. However, he had a reputation for alcoholism and promiscuity, and he embraced western hippie ideology.[33],[34] He justified his lifestyle as "crazy wisdom," and he claimed that he was able to convert alcohol into water inside his body, but unfortunately, his tragic early death was attributed to chronic liver disease caused by heavy alcohol use.[35],[36],[37] His heir, Ösel Tendzin, heavily influenced by his philosophy, later contracted HIV but kept silent and sexually active for at least two years.[38] We found that English search results about Chögyam Trungpa tended to downplay his promiscuous lifestyle and primarily focused on his books, teachings, and meditation centers, but Chinese search results had lengthy articles detailing his notorious lifestyle.

---

[4] Buddhism in Vietnam and Thailand primarily follow Theravada Buddhism, which Pali scriptures like the Jataka tales (stories of past incarnations of the Gautama Buddha) are important parts of Theravada Buddhism. Unlike a creator God, the historical Buddha is portrayed as a human who has many past lives and attained enlightenment through good karma and meditation.

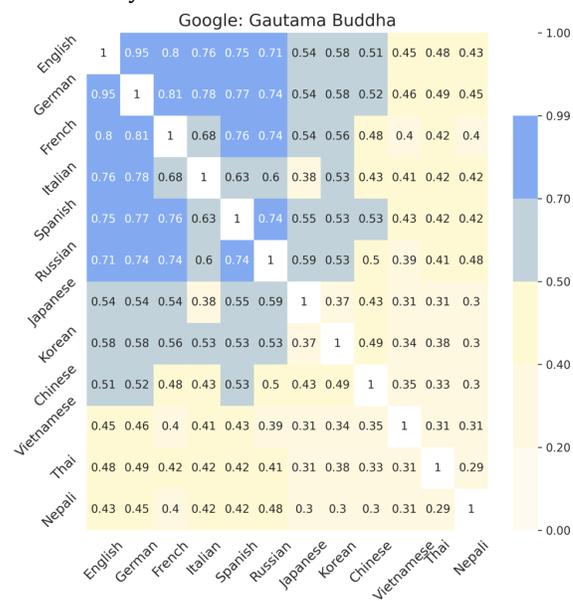

**Figure 7** This figure shows the TF-IDF cosine similarity scores for Google's top 50 website titles on "Gautama Buddha." It presents a cluster between European languages around the threshold of 0.7.

The current 14th Dalai Lama is well known worldwide as a Nobel Peace Prize laureate. Japanese, French, and German contents were the most similar to English language content, Chinese had the least similarity with it. The site nobelprize.org had a lengthy biography about the Dalai Lama, and it always appeared among the top 5 on Google using English, French, and German. This website was, however, invisible in Chinese search results because this website was not available in Chinese. Thus, Chinese users were completely oblivious to the Dalai Lama's biography on nobelprize.org, and their search results primarily focused on his political controversies with the Chinese government. Figure. 8 shows that using the search term "Dalai Lama," Chinese search results contained a large portion of websites ending with .gov.cn, .com.cn, and .cn which were regulated by the Chinese government. We found that messages on government-controlled and non-controlled websites presented polarized views about the Dalai Lama. Some condemned him for the self-immolation incidents in Tibet in protests against "China's occupation of Tibet," while others addressed his insistence on compassion and peace despite China's actions toward Tibetans.

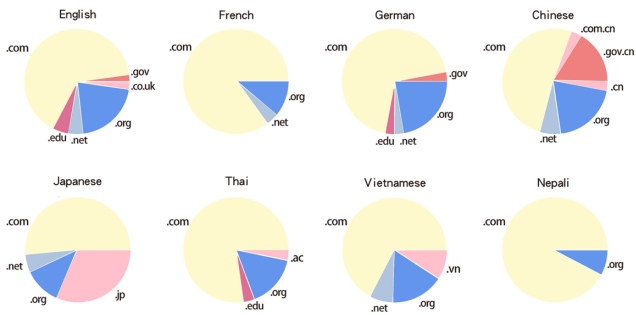

**Figure 8 This figure shows website domain distributions for Google's top 50 website titles on "Dalai Lama." A substantial percentage of Chinese-language websites are hosted under the .gov.cn domain.**

In this case, by scanning the English and Chinese search results about Chögyam Trungpa and the Dalai Lama, a reader can easily conclude that the English version was philosophical, peaceful, and institutional, while the Chinese version was scandalous, dark, and censored. This can be dangerous to both language communities. To English users, being completely oblivious to the dark side of Buddhism will trap them into a beautiful illusion that makes them vulnerable to unethically intended religious leaders or organizations. On the other hand, to Chinese users, being inculcated with overtly negative news about Buddhism will deprive them of the opportunity to engage with the wholesome practices of Buddhism.

We find that language bias becomes more pronounced for narrower topics. In other words, the portion of the elephant we can perceive on Google gets smaller for narrower topics. The four examples discussed above – "Karma in Buddhism," "Gautama Buddha," "Chögyam Trungpa," and the "14th Dalai Lama" – further highlight Google's perpetuation of cultural stereotypes through its ranking algorithm. The causes of this heightened level of bias might be related to the lack of diversity in terms of the numbers of available websites on narrower topics, and in turn, leading to the distribution of the top-ranked websites more concentrated around a fewer number of subjects.

## 3.2 Buddhism-related queries on ChatGPT

ChatGPT is considered a new wave of AI technology that could replace traditional search engines like Google. We discuss both ChatGPT-OpenAI and ChatGPT-Bing in this section.

First, we experimented with ChatGPT-OpenAI with some Buddhism-related questions. We found that even though it took many languages as inputs and outputs in the input language, it 'thought' in English. We asked ChatGPT-OpenAI "who is Amitabha Buddha" in multiple input languages, and we repeated this experiment several times. Its responses were almost identical regardless of the input language. In particular, its answers contextualized Amitabha Buddha within the Pure Land tradition of Mahayana Buddhism, which was very similar to Google's top search results in English, whereas the prominent features in the Chinese search like "how to chant 'Namo Amitabha Buddha'" were absent in its responses. Likewise, we asked ChatGPT-OpenAI "give me a Buddhist story of karma" in multiple input languages. The generated karmic stories were fake and did not contain distinct cultural elements except for the character's name (e.g., an English name for English input or a Chinese name for Chinese input). For example, when we used Chinese as the input language, the distinct Chinese cultural elements were absent, such as Confucian ideas or Chinese historical contexts, whereas they showed up prominently on Google and YouTube. However, we also noticed that ChatGPT-OpenAI was able to write in Classical Chinese and construct fake Classical Chinese quotes, indicating it understood Classical Chinese syntax and was able to 'think' in Classical Chinese to some extent. As it was primarily trained on English language data, the English perspective still overtly dominated its 'internal logic.'

Next, we conducted similar experiments on ChatGPT-Bing. Like Google Search, ChatGPT-Bing looked within the input language's repository by default, i.e., it searched within French websites for an input question written in French. We noticed that its responses present nuanced modifications across various input languages – they tend to differ only by *a few* words but convey different perspectives and messages. For example, ChatGPT-Bing's answer to "what is Buddhism" described Buddhism as "a religion and philosophy" in English but as "an atheist religion" in Thai.

With the interface of ChatGPT, the direct connection between the user and the 'elephant' is lost. ChatGPT, acting as a cyberspace delegate, is heavily reliant on English sources. As it turns out, we almost get one single unified Anglo-American perspective through ChatGPT. This is harmful to users of all languages. English users are caged inside their own cultural realms, rarely breaking cultural boundaries to encounter diverse ideas, while non-English speakers are often unknowingly imbued with Anglo-American viewpoints, with their native perspectives being systematically undervalued without scrutiny.

## 3.3 Buddhism-related videos on YouTube

In general, Google, an indexed hodgepodge of all resources online, is the primary place for us to start a search. YouTube is a frequently returned platform on Google and a popular site for Buddhist practitioners to share music and meditation tutorials. In this section, we provide analyses on YouTube to examine whether language bias exists and how it differs from or is similar to that of Google.

*3.3.1 "Buddhism" videos on YouTube*

Like our analyses on Google, we started with the term "Buddhism" on YouTube. As Figure 9 demonstrates YouTube's TF-IDF cosine similarity scores show that there is *little* similarity across languages and lack apparent clusters. Compared to that of Google, we find that YouTube's language bias is more pronounced and filtered by the now-trending events and popular YouTubers channels. They tended to zoom into the details of a very small concept, such as instructions on breathing meditation and recordings of Buddhist festivals. Compared to text-based information on Google, a typical YouTube user can hardly get a broad picture of Buddhism. YouTube videos provided a more ethnocentric experience due to the visual and auditory presentation of the host's face and voice.

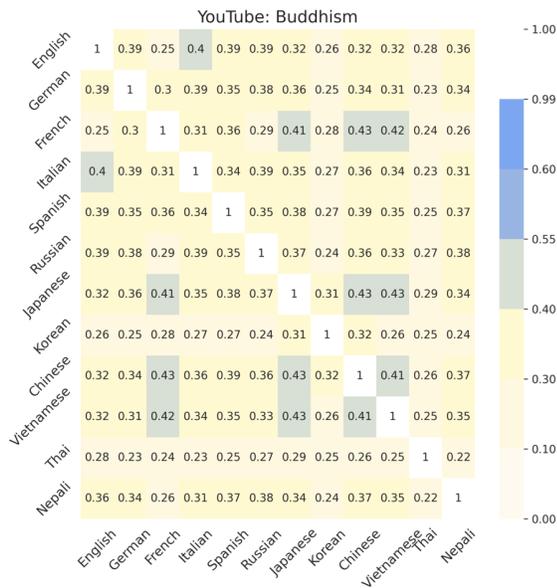

**Figure 9 This figure shows the TF-IDF cosine similarity scores for YouTube's top 35 video titles on "Buddhism." The data reveals no discernible clustering pattern.**

We found that the top-ranked videos on YouTube are *highly diverse* in general. Despite this general diversity, some of the top-ranked videos in European languages centered on the theme of 'how to deal with adversity in life from the wisdom of the Buddha' (though their exact titles vary), which Asian language videos differ dramatically in themes. For example, English videos had Indian, Tibetan, and American Buddhist monks teaching how to calm the mind; French videos had interviews of Caucasian French monks and the Buddha's wisdom for going through difficult times; and German videos had life lessons from the Buddha. Comparatively, Japanese search results had interviews with Japanese Buddhist scholars and a video of a Japanese monk performing Buddhist music. There was a top-ranked video titled "Easy-to-understand 15-minute explanation of the differences between Buddhist schools" that gave a brief overview of the major Buddhist schools, but it should have been titled 'differences between *Japanese* Buddhist schools' because it *only* compared Japanese Buddhist schools.[39] Vietnamese search results had several recordings of grassroots volunteer activities organized by Vietnamese non-profit Buddhist organizations. Tibetan search results were uniformly Tibetan lamas preaching Buddhist doctrine. Chinese search results had lecture recordings on the *Diamond Sutra* and videos by a popular Chinese talk show host, Yuan Tengfei, who has over 447K subscribers and happened to have episodes on Buddhism.

We also observed that YouTube videos, except in English, were primarily produced by the dominant ethnic group in their language community, i.e. French videos were hosted by French people and Vietnamese videos were hosted by Vietnamese people. But English videos were commonly hosted by non-native English speakers. It is intriguing to see that when non-native English speakers host English videos, they tended to talk about stuff that interests the English community rather than their native ethnic group. For example, the Tibetan monk Yongey Mingyur Rinpoche had attracted 546K subscribers and had several English videos ranked highly under the English search term "Buddhism." In his English videos, he mainly talked about life wisdom and philosophy from the Buddha – things that follow the mainstream English interests, rather than how to properly chant a mantra or receive blessings from the Buddha – things Tibetans do daily.

Based on our observation, YouTube search is similar to Google search, where they rank websites/videos based on some weighting factors, in contrast to Wikipedia, which directs users to a single article. YouTube's ranking system seems largely determined by the freshness, or recency, of the video, as well as its popularity. This popularity is reflected in the number of views, likes, and comments a video receives and the number of subscribers to the YouTube channel hosting the video. This popularity also seems to be related to users' interest and satisfaction. Videos that align with more people's interests or preexisting cultural beliefs receive more likes and subscribers and are promoted through the recommendation section on YouTube.

### 3.3.2 Popular narrower topics related to Buddhism on YouTube

Like our analyses on Google, we narrowed down our search topic to popular subtopics in Buddhism. Our YouTube findings are similar to those we found with Google: European language search results tended to ask 'what it is,' while Asian language search results emphasized 'what to do' for belief-oriented terms. On the term "Karma in Buddhism," European language videos emphasized the definition and basic understanding of karma, while Asian language videos explained karma through stories. However, these stories varied dramatically by language. For example, Chinese videos talked about karmic stories of husband and wife that drew on the imperial examination (科举) from the Chinese historical context.[40]

English videos on the search term "Dalai Lama" tended to emphasize his wisdom and approach to happiness while largely ignoring his political background and conflicts with the Chinese government. In contrast, Chinese videos uniformly focused on his political issues with the Chinese government. Among the top 20 ranked videos, not a single video focused on his wisdom and life lessons but contextualized him within political affairs. Thai search results had interviews and teachings of the Dalai Lama. Similar to the Tibetan monk Yongey Mingyur Rinpoche's video discussed above, the Dalai Lama talked about loving-kindness, a common doctrine in both Thai and Tibetan Buddhism, to Thai followers, rather than Tibetan tantric visualization technique, which is not widely known in Thai Buddhism.

Language bias on YouTube resembles that of Google but is more pronounced. The top-ranked YouTube videos followed the dominant Buddhist traditions in different language communities, i.e., Japanese videos mainly talked about Japanese Buddhism, and European language videos followed the western Tibetan Buddhist tradition. This dynamic use of audio and visual information creates a deeper connection between viewers and the content than Google's textual information, constructing a more ethnocentric experience for the viewers. In that respect, YouTube videos zoom in to the 'color' or 'texture' of the elephant, but the 'color' the audiences see or 'texture' they feel is different based on their search languages.

### 3.4 Buddhism-related articles on Wikipedia

While investigating the top-ranked websites on Google, we noticed that as long as a relevant Wikipedia article existed, it was always ranked highly across all languages. Due to Wikipedia's prominent position and popularity, it equally plays a dominant role in shaping public opinion. In this section, we provide language bias analyses on Wikipedia and discuss the extent to which language bias exists on Wikipedia, and how it compares to Google's.

#### 3.4.1 "Buddhism" on Wikipedia

Like our analyses on Google, we started with the term "Buddhism" on Wikipedia. Compared to Google's language bias, we found that Wikipedia articles' content titles mainly differed in topic coverage but not much in tones of voice. The preferences of topics tended to correlate with the dominant intellectual traditions and camps in different language communities.

On the search term "Buddhism," the French page started its article by defining Buddhism as "a religion and a philosophy."[41] It had a long section that engaged with the major debates of Buddhism, such as whether Buddhism is a religion or philosophy, or both. But such debate did not take an equivalent space in the English or Asian languages' pages. In fact, this topic is one of the central scholarly debates in French academia, with the general argument being that –

Buddhism does not fit the Western characterization of "religion" because Buddhism is not based on divine revelation but also does not fit the definition of "philosophy" because its doctrine aims towards the ultimate goal of salvation rather than universal reasoning.

'Religion' is a more dispersed idea in America, where it could refer to a diverse set of spiritual beliefs and practices. Comparatively, 'religion' is mostly associated with the catholic church in France, and France has a rich intellectual tradition in philosophy – figures like Michel Foucault, Jacques Derrida, and Jacques Lacan used to give philosophical talks on TV. When French scholars approach Buddhism, it might be natural for them to translate their dominant intellectual frameworks of religion and philosophy to Buddhism, attempting to categorize Buddhism into their definitions of religion or philosophy. However, the debate of "whether Buddhism is a religion or philosophy" is *not* unique to French academia. English and German scholars do write on this subject, but their Wikipedia pages do not have a comparably detailed treatment. This might have been due to an incomplete coverage of what's important in a language community on Wikipedia.

Looking at the Chinese article on "Buddhism," it addressed topics like "dharma name," "cloth," and "hairstyle" that did not exist on other languages' pages. There are several potential causes for its special treatment on these issues. First, many Buddhist texts, such as the *Lankavatara Sutra* (楞伽经) and *Vinaya Piṭaka* (律藏), that address these issues were translated into Classical Chinese during medieval China, and these texts are still widely circulated in China today. Second, according to the official central Chinese government statistics, there are over 33,000 monasteries in China,[42] so people who are interested in writing Wikipedia articles might think it is helpful to address these issues on Wikipedia. However, like the pattern in the French article, Vietnam, Thailand, and Nepal all have millions of Buddhist practitioners, and the *Lankavatara Sutra* and *Vinaya Piṭaka* are also widely circulated among South Asian Buddhist traditions, but their Wikipedia pages do not address these issues like the Chinese article.

Likewise, the Vietnamese page had detailed accounts of Buddhism in many South Asian countries like Vietnam, Burma, Sri Lanka, Cambodia, Thailand, and Laos. The appearance of other South Asian Buddhist traditions on the Vietnamese page feels natural because they are geographically closer and the majority of Buddhists in these countries follow the Theravada Buddhist tradition. In contrast, the Thai and Burmese articles did not have similar sections.

For such a broad topic like "Buddhism," it appears Wikipedia articles correlate with cultural significance and the intellectual circles in its language community as they tend to emphasize subjects that are important to their own communities. However, it is important to note that as Wikipedia constantly undergoes refinements and updates, its topic coverage in a given time frame might be incomplete to fully represent what's significant in a language community.

### 3.4.2 Narrower academic research topics related to Buddhism on Wikipedia

To further examine the patterns and causes of variations in topic coverage on Wikipedia, we narrowed down our search to some central Buddhism-related topics like "Buddhist meditation," "Gautama Buddha" and "Buddhist philosophy."

Like Google, language bias increased for narrower topics on Wikipedia. On the term "Buddhist philosophy," the German article compared Buddhist philosophy to Western philosophy, primarily citing German scholars like Schopenhauer, Heidegger, and Nietzsche. In fact, these 19th-century German intellectuals brought Buddhism into Western philosophy. Schopenhauer's central philosophical argument was influenced by Indian thought, particularly the ideas found in the Upanishads, Buddhism, and Vedanta philosophy. His central philosophical argument revolves around the concept of the "will," which he considers the ultimate metaphysical reality behind all existence.[5] Heidegger's writings have many parallels with Zen Buddhism philosophy,[6] and Nietzsche saw Buddhism as a nihilistic religion, and his view became very influential in the West.[7] The French article also has a section on Western philosophy, but it compares Buddhist philosophy with Ancient Greek and modern European philosophy. There is a tradition in French intellectual circles of comparing Asian philosophy, particularly Buddhist philosophy with ancient Greek philosophy and modern European philosophy. This tradition began in the late 19th and early 20th centuries. It was led by scholars including Paul Demiéville, André Bareau, Étienne Lamotte, and Jean Przyluski, who were prominent members of the French School of Asian Studies. These scholars contributed significantly to studying Buddhism and its relationship with Western thought, particularly through their comparative studies of Buddhist and Greek philosophy. They also paved the way for future scholars to continue this line of inquiry. On the other hand, the Chinese, Japanese, and

Vietnamese articles are very short on Buddhist philosophy. They do not compare it with Western philosophy because they might view Buddhism as natural to them and do not see an apparent and direct connection between Buddhism and Western philosophy.

The English article on "Buddhist meditation" contained a lengthy introduction covering all major Buddhist traditions, including early Buddhist schools, Indian Buddhism, East Asian Buddhism, Tibetan Buddhism, etc. But the Chinese Wikipedia article only talked about mediation within the realm of Chinese Buddhism. The article's overall writing style – patched with lengthy quotations from Buddhist scriptures written in Classical Chinese, was very similar to the way most Chinese Buddhist abbots preach meditation in monasteries, where they recite passages from the original Buddhist texts written in Classical Chinese. The Vietnamese article strongly emphasized mindfulness meditation, which is more in line with spiritual training in Buddhist chaplaincy rather than rigorous academic scholarship. The focus on mindfulness meditation is a common practice in most Buddhist organizations. On the other hand, the French article contained a "Buddhist meditation and neuroscience" section, which did not exist in any other language contents. In fact, "Buddhism and neuroscience" is a popular topic among the French Buddhist community. The well-known "World's Happiest Man" measured by brain gamma waves is the French monk Matthieu Ricard who has done a series of neuroscience experiments to prove the benefits of meditation. The Nepali article was very short and the German and Thai Wikipedia articles did not have pages available on this term.

Similarly, on the term "Gautama Buddha," the German article had a lengthy section on "chronology" about dating the historical Buddha, but the English one only had a short paragraph on chronology, and this concept was absent in other languages. In fact, German scholarship played the most significant role on the chronology of dating the historical Buddha, and this intellectual tradition came from their scholarships on dating the historical Jesus. German academic institutions have a philologically oriented tradition based on primary texts. The studies on dating the historical Buddha heavily rely on close readings of ancient manuscripts written in Sanskrit, Pali, and Tibetan to trace out evidence of early records of the Buddha. The *Dating of the Historical Buddha* edited by Heinz Bechert, a German Indologist and Buddhologist, is a two-volume collection summarizing the major issues and debates on this topic. This two-volume collection was compiled after the 1988 symposium on "The Date of the Historical Buddha and the Importance of its Determination for Historiography and World History" held under the Akademie der Wissenschaften in Göttingen (Germany). Many contributors in this work have a German intellectual background, and a large portion of their cited works are written in German.

Language bias becomes more pronounced for narrower topics on Wikipedia. As demonstrated above, the available intellectual repositories, i.e. books, scriptures, and scientific experiments, on "Buddhist meditation" and "Buddhist philosophy" across Asian and European languages were very different. And Wikipedia contributors rarely cited sources across languages, i.e. the English article primarily cited English books, articles, news or other sources, and the French article cited French ones, and so on. This made topic coverage across languages highly correlated with the extant intellectual database in different language communities. It is important to note that figures who are able to influence the intellectual circles might not necessarily be university professors or scholars like Nietzsche, but who have intellectual power over a large crowd, like the Dalai Lama, a monastic abbot, popular politician or organization.

---

[5] He draws parallels between Buddhist philosophy and his own ideas about the nature of reality and suffering, and the path to enlightenment in *The World as Will and Representation* and *The World as Will and Representation*. He also provides a detailed analysis of Buddhist concepts such as the Four Noble Truths and the Eightfold Path, and sees Buddhism as a pessimistic philosophy that offers a way of coping with hardships.

[6] The conceptual similarity and difference between Heidegger and Zen Buddhism has been widely studied in philosophy. Both Zen and Heidegger believe that metaphysics must be transcended, and they agree that a radical transformation of consciousness is necessary to transcend the dichotomous matrix of ordinary thinking. Both of them aim to disclose primordial experience by dismantling the sedimentations involved in language and conceptual thinking.

[7] Nietzsche believed in self-overcoming and the internalization of the will to power, while Buddhism promotes overcoming inner forces through mind development. Both systems reject the idea of an eternal substance or God and search for a solution to the meaninglessness of life by turning inward and advocating a transformation of inner forces and outlook on the world.

Unlike Google's language bias that is filtered by both topic coverage and cultural-political views, Wikipedia articles mainly differ in topic coverage but not much in views. Compared to Google, which contains both first-person and third-person content, Wikipedia articles contextualize first person narratives as third-person observations. Information is written in a descriptive tone and free of arguments and persuasion. In a sense, Wikipedia presents a silhouette of the elephant, that the 'color' and 'texture' of the elephant are missing.

## 3.5 Complex topics that have not been strongly influenced by language bias

We explored a wide range of complex topics and found that it was not easy to find terms that did not exhibit language bias. However, we did observe some specific technical or mathematical terms like "automatic differentiation," "Jacobian matrix," and "Hidden Markov chain" that were not strongly influenced by language bias.

We identified three unique traits of these terms. First, they have a clear and fixed definition independent of the culture they are born with. Second, they are known only by a small group of people that the vast majority do not know or care about. Thus, any discussion or writing about these terms is typically created by individuals with knowledge and expertise. And third, they have a short history. Scientific terms like "Newton's first law," "quantum entanglement," and "solar system" that have a longer history are often tangled with historical and cultural discussions, and thus articles written on them are inevitably filtered by language bias. For example, on the search term "solar system," Google's English search results uniformly returned websites writing about the solar system within the realm of astronomy. But East Asian language search results contained lots of non-scientific websites. Chinese results talked about one's childhood dream about the solar system; Vietnamese results contained websites about the "Solar powered water system" and "solar panel." In this case, the dominance of scientific descriptions of the solar system in English search results might be because the English repository has a large number of scientific articles because researchers tend to publish scientific papers in English.

## 4  SOCIAL IMPLICATIONS

Throughout its 2,500-year history, Buddhism has spread from its origins in India to nearly every Asian country, and it has also made its way to the West in recent times. As a result, a rich and diverse Buddhist tradition has blossomed worldwide. However, English users get a tiny piece of this tradition online, with a strong focus on topics like meditation and life wisdom. Aspects highlighted in Chinese or Vietnamese Buddhism are not readily visible in English searches, while the English's emphases on meditation and life wisdom are not prominent in searches conducted in other languages. Due to language bias, accessing information from other cultural perspectives can be challenging for all language users. It becomes natural for us to evaluate people from Buddhist traditions based on our own cultural perception of Buddhism.

In a similar vein, liberalism emerged as a prominent political ideology largely owing to the contributions of British and American thinkers. When it was exported to the rest of the world, it underwent various transformations as it adapted to diverse local contexts and political environments. Like the problem we witness with Buddhism, each language community gets their culturally dominant views of liberalism on various platforms like Google and YouTube and the singular Anglo-American perspective on ChatGPT. Since liberalism is the founding ideology in the U.S. and U.K., our investigations revealed very little *criticism* about liberalism in the top English search results on Google and exactly no references to neo-liberalism.[8] However, French, German, Italian, and Spanish search results *all* contained a few articles discussing the limits or downsides of liberalism – "Liberalism: Do away with clichés and misconceptions"[43] in French, "The Crisis of Liberalism in the Interwar Period" in German, "Tolerance and the limits of liberalism." in Italian, and "Why 'Western' Liberalism Won't Save Latin America" in Spanish.

The endorsement of liberalism became more pronounced in English when we inquired on Google, "Is liberalism good?" We got answers from *The Economist*, a British weekly newspaper, acclaiming "Liberalism is the most successful idea of the past 400 years"[44] and *The American Prospect* explaining "Why Liberalism Works."[45] Even with the search term "drawbacks of liberalism" in English, over half of the articles returned tend to compare *both* its advantages and disadvantages rather than solely emphasizing its drawbacks. The U.S. and U.K. embrace liberalism because it defines the guiding principles for their respective political systems. French, German, Italian, and Spanish societies are deeply influenced by liberalism, but the dominant liberal theorists are British thinkers like Thomas Hobbes, John Locke, and John Trenchard. We found that Google's top search results do present slightly different views in these non-English language search results, and the roots of liberalism might partially explain these findings.

---

[8] On the search term "Liberalism." Accessed March 1, 2023.

On the other end of the spectrum, virtually all Asian languages we tested returned one or more websites that talk about 'neoliberalism,' which has a pejorative connotation – "Neo-Liberalism – Doimoi.org" in Vietnamese, "We don't need 'new' liberalism, we already have liberalism" in Korean, "Why British industry has declined and hollowed out, and what role has neoliberalism played?" in Chinese. The pushbacks of liberalism in many Asian countries can be attributed to the fact that the emphasis on freedom introduces a set of moral norms and behaviors that discount well-established social and cultural values, such as importance of family responsibility and national unity.

We also observed that different language websites offer varying definitions of liberalism on Google. Top English sites described "liberalism" as protecting individual rights, liberty, freedom, and equality, enabling private companies to thrive with minimal government interference. In contrast, the Vietnamese article "Liberalism as a radical centrist" by Clark Ruper defines it as "an ideology that doubts and challenges the use of political power."[46] As we transitioned to ChatGPT-Bing, we asked it "What is liberalism?" in English, Chinese and Vietnamese. Its definitions were closely aligned with Google's top search results. In English, it defined liberalism as "a political and moral philosophy."[9] In Chinese, liberalism was described as "an ideology and philosophy with freedom as its primary political value" and emphasized the "limitation of government power."[10] In Vietnamese, it was characterized as "a political and philosophical ideology" that "opposes state intervention in personal and economic life."[11] On the other hand, ChatGPT-OpenAI's definitions of liberalism were confined to a singular Anglo-American perspective, emphasizing "individual liberty," "equality," and "freedom" regardless of input language. Aspects like 'questioning political power' and 'limiting government interference' found in Chinese and Vietnamese sites were absent.

Critics in Asian countries often argue that liberalism's emphasis on individual rights and freedoms in fact exacerbate economic inequality. The current U.S. political economy of liberalism has been increasingly questioned due to its failure in delivering effective solutions to deal with many of the pressing issues, such as the significant increase in income inequality and limited upward social mobility. Liberalism's failure to address income inequality is rooted in its core principles of free market, deregulation, and privatization. The unchecked pursuit of these principles led to the collapse of Lehman Brothers in September 2008 as an aftermath of the neoliberal policy of deregulation, which allowed large investment banks to take excessive risks. When some people realized that 'free market' and 'self-autonomy' were being pushed to an extreme, neoliberalists criticized social policies designed to help people escape difficult circumstances. They claimed that they would not work if the state paid people not to work. Under the guise of 'free market' and 'self-autonomy,' powerful individuals and corporations grow strong enough to sway government decisions. With the fall of the Berlin Wall in 1989, liberalism seemed to have served as the normative model for the rest of the world to follow. However, the successful establishment of liberalism and democracy relies on premises like firmly established law and media regulation to curb the influence of powerful elites, preventing propaganda and manipulation of decision-making. Since English speakers cannot access information from local perspectives of other countries, it is easy for them to overlook local conditions and judge their policies through an ethnocentric lens, disregarding the differences in various state enforcement capabilities. This can potentially aggravate the sufferings of individuals in those regions by imposing a Western-centric policy, which may not fit their unique circumstances and needs. Sadly, the invisible yet pervasive algorithmic force has only fed our egos by reaffirming our cultural beliefs and silencing views outside our own.

Another critique from Asian countries is that liberalism can result in *moral decay* because they see excessive "rights" and "freedom" can slide into a permissive and morally ungrounded stance. This perception is reinforced by extensive news coverage on issues like drug use and gun ownership in America, often defended under the banner of "freedom." Provocative themes such as drug consumption, gender identity, sexuality, and violence are promoted in American pop music and Hollywood films, which are the primary sources for people living in Asian countries to directly

---

[9] Liberalism is a political and moral philosophy based on the rights of the individual, liberty, consent of the governed, political equality and equality before the law [1]. It emerged from the Western Enlightenment in the 16th and 17th centuries [1][2]. It has different types, such as classical liberalism, social liberalism, neoliberalism, etc. What type of liberalism are you interested in? Sources: [1] en.wikipedia.com [2] britannica.com [3] bing.com. Retrieved March 1, 2023.

[10] The original output in Chinese: 自由主义是一种以自由为主要政治价值的意识形态和哲学 [1]。它主张保护个人的思想自由、权利、财产和平等，以及限制政府的权力 [2][3]。自由主义起源于 16 世纪欧洲的宗教战争，反对君权神授、世袭制和国教制度 [1]。自由主义有不同的类型，例如古典自由主义、社会自由主义、新自由主义等。你对哪种类型的自由主义感兴趣？Sources: [1] zh.wikipedia.org [2] baike.baidu.com [3] zhuanlan.zhihu.com [4] wikiwand.com. Retrieved March 1, 2023.

[11] The original output in Vietnamese: Chủ nghĩa tự do là một hệ tư tưởng chính trị và triết học dựa trên các giá trị về tự do và bình đẳng [1][2]. Nó phản đối sự can thiệp của nhà nước vào đời sống cá nhân và kinh tế [3][2]. Nó bắt nguồn từ phong trào Khai sáng ở phương Tây vào thế kỷ 16 và 17 [1][2]. Bạn muốn biết về lịch sử hay các loại của chủ nghĩa tự do? Sources: [1] vi.wikipedia.org [2] luatminhkhue.vn [3]

bing.com [4] chungta.com [5] vi.wikipedia.org. Retrieved March 1, 2023. Retrieved March 1, 2023.

experience American culture. By directly accessing information from English-language websites, Asian language speakers and other non-English users may discover that American life encompasses far more than what is portrayed in newspapers, music videos, and films.

However, the additional benefits non-English speakers might gain from culturally-embedded ideas through exploring English websites could be limited compared to native English speakers. Due to the global prominence of the English language and the U.S. dominance in global affairs, British and American culture have been widely exported to the rest of the world. For non-English speakers, English culture and values have been largely woven into their quotidian through textbooks, entertainments, and news either as sources of knowledge or issues of conflict. In contrast, English speakers are not naturally exposed to other cultures unless they actively choose to study them. Media coverage of non-English cultures often focuses on negative stories like natural disasters or controversial events, rather than exploring the complexities and values of ordinary life. Therefore, exposure to diverse cultures is crucial for English speakers to form a more nuanced understanding of the world.

# 5 CONCLUSION

In conclusion, we find that top search results across a range of platforms promote the dominant cultural phenomena based on the language of a search query. The dominant cultural phenomena can relate to many factors, such as the majority's interests, a major social event like the presidential campaign, or a powerful agency's agenda. Even on Wikipedia, articles reflect the dominant intellectual traditions or trends in a language community. The dominant intellectual trends are not necessarily tied to scholars in academia but to people who have intellectual power over a large crowd, like a popular politician. User-oriented ranking algorithms like YouTube's that rank on the freshness and popularity of a video are more likely to gain popularity by following the majority's interests and their pre-existing beliefs, such as teaching a Westerner about the Buddha's wisdom in dealing with life adversity rather than how to properly chant "Amitabha Buddha." The problem with language bias is compounded by ChatGPT. As it is primarily trained on English language data, it presents the Anglo-American perspective as truth – as if it were the only valid knowledge. In addition to pure language models, we also observe that language bias is intertwined with multimodal AI models that take textual input. In text-to-image models like DALL-E, the generated images look significantly different in terms of style and content across various language inputs.

We have described that Google presents a piece of the elephant based on a user's query language. Its top-ranked search results vary in both topic coverage and views. Wikipedia shows a rough silhouette of the elephant, but the silhouette – topic coverage – differs by language. It writes in a descriptive tone and contextualizes first-person narratives and subjective opinions as cultural, historical, or religious phenomena. YouTube displays the 'color' and 'texture' of the elephant as it incorporates images and sounds that are effective in evoking emotions. Since its ranking is closely tied to popularity, the top-ranked videos tend to create a more profound ethnocentric experience as they zoom into a highly confined range of topics or views that conform to the majority's interests. ChatGPT, covered itself as a multilingual 'sighted' delegate of Search, in fact is mostly 'blind' to non-English perspectives, constructing a delusional experience of authoritativeness to users.

The issues surrounding language bias on digital platforms have largely remained hidden in the past. Many computer scientists seem to have assumed that reflecting "common case"[47] or aligning with the "majority's view" within their own culture, which naturally aligns with their profit-seeking intent, is the correct approach. On the contrary, this in fact becomes the 'tyranny of opinion' and 'tyranny of the majority,' discounting the rights of the minority, fostering intolerance among different cultural-linguistic groups, and posing an alarming threat to our cherished ideas of democracy, free speech, human rights, and equality. Individuals' exposure to "diverse and antagonistic views"[48] as phrased by the Supreme Court is essential in cultivating well-informed and self-reflective citizens within a democratic society where debate isn't dominated by corporations, politicians, or privileged groups.

The way algorithms have been run has trapped us inside a self-perpetuating cycle. Platforms like Google, ChatGPT, and YouTube pick up the dominant views within a language corpus and circulate this information to the public. Unpopular or counter views are presumably considered as noise by the system. The public absorbs and reinforces these ideas, which are then picked up by the system once more, perpetuating the cycle. While platforms like Wikipedia operate with a separate algorithmic framework, its users and readers are embedded within this loop, causing its content to also revolve within this cycle. As if we were agents stuck at a local minimum in a machine learning environment, where algorithms continue to sample information inside this local minimum and discourage exploration beyond its boundaries. Being entrenched in this self-perpetuating cycle limits our capacity for growth and critical thinking. Just as each of the blind person believes what they see is the totality of truth, we never encounter our own 'blind spots' without exposure to diverse and even opposing views.

Computer scientists may excel in programming languages, but this expertise does not imply that they have comprehensive knowledge across all domains. However, their technology impacts every domain and individual in society. It is essential to involve professionals from diverse fields, particularly humanities, social sciences and government, to collaboratively examine the broader social and political consequences of emerging technologies.

We hope our study can alert the general internet users of the implicit biases they consume every day and inspire future scholars to continue investigating the larger socio-political impact of new technologies: If we see a search engine or an AI bot as forms of new media that transmit information, should we expect them to obey similar moral principles and political responsibilities as traditional media outlets like newspapers and radio stations? Is Section 230 outdated? Alternatively, can we view the internet as a public infrastructure as it provides free access to information and facilitates the functioning of a civilized society? Should we then expect Google and similar platforms to adhere to certain political obligations?

Given the current state of AI architectures like the Transformer, it is hard to see the existential threat posed by AI is nearly possible. However, it is much more likely that AI would cause humans to turn us against each other by telling us different stories. If we continue this way, we would expect different linguistic groups to be even more divided in a decade than they are today while steadfastly holding the conviction in possessing the ultimate truth.

## ACKNOWLEDGEMENT

The paper being presented was developed from the course project from AC221: Critical Thinking in Data Science taught by Michael D. Smith in Spring 2022 at Harvard University. We would like to express our sincere appreciation to Xiaohan Yang, Xingyu Liu, and Wenqi Chen for their initial contributions to this paper as a course project. We would also like to thank Peter Bol, David Atherton, Hoa Le, Hanrui Wang, Wenfei Wang, Changzai Pan, and anonymous reviewers for their valuable comments to this paper.

# Appendix 1 Previous Work

Google search engine bias has attracted scholars' attention in recent years. In particular, a number of scholars have examined political bias embedded in search engine systems. Juhi et al. (2019) propose a framework to quantify political bias in the search system and discern the source (input data or ranking system) of bias.[49] A. Diaz (2008) investigates socio-political bias on Google search and concludes that Google may favor popularity over richness.[50] Wallis (2011) also points out that the search results on Google are inclined to the monopoly of culture and knowledge of the dominant communities.[51] Noble (2018) examined how Google search results can enhance stereotypes and racial biases.[52] These studies raised compelling evidence and arguments about how search engine bias influences user's opinions and perceptions, but little work has been done on language bias in search engines. Christofol et. al (2021) investigated language bias in Google Scholar ranking system and found that articles written in languages other than English are being systematically relegated to positions that were virtually invisible to the public.[53] Published in 2021, Christofol et. al's work shed light on language bias and raised the issue that less popular languages may be put in disadvantageous positions due to underrepresentation in search engines.

Rozad (2023) investigates the potential political biases of Large Language Models (LLMs) like OpenAI's ChatGPT, given their potential commercial applications and role as gateways to human knowledge.[54] The author conducted 15 political orientation tests (14 in English, 1 in Spanish) on ChatGPT and found that 14 out of 15 tests showed a preference for left-leaning viewpoints. Despite ChatGPT often claiming to hold no political opinions, the results suggest a potential bias somewhere in the construction of the system. Kirdemir et. al (2021) examine the presence of political bias in YouTube's machine learning recommendation algorithms, specifically within its search and video recommendation systems.[55] The findings revealed that while the average bias of highly-ranked videos leans slightly Democratic, YouTube's "Up-Next" recommendations tend to minimize the magnitude of bias. Lutz et. al (2021) examine the structure of recommendation networks and probabilistic distributions of the node-centric influence of recommended videos, using a stochastic approach and analyzing 803,210 recommendations based on search queries and seed datasets from previous studies.[56] Their study reveals a systemic tendency for YouTube's video recommendation system to impose bias in favor of a tiny fraction of videos in each scenario.

Besides Christofol et al.'s research on language bias in the Google Scholar ranking algorithm, research on algorithmic language bias is limited. Studies that have touched on issues involving language bias typically focus on tasks like detecting different cultural perspectives in multilingual datasets using NLP methods or identifying cultural-language correlation, rather than seeing them as a type of algorithmic bias tied to search language. Tian et. al[57] study colingual groups and use language corpora to identify and measure distributional perspective difference using NLP methods based on a collection of Wikipedia pages for English, Chinese and Japanese. Góngora-Goloubintseff[58] focuses on the challenges of maintaining a Neutral Point of View (NPOV) in Wikipedia articles by examining visual and textual data from English and Spanish Wikipedia entries on the Falklands/Malvinas War. Góngora-Goloubintseff reveals that certain images and lexemes in specific contexts can indicate NPOV violations. Wolniewicz-Slomka[59] examines three Holocaust articles on Wikipedia in English, Hebrew, and Polish, based on the assumption that collective memory differences are manifested on popular knowledge platforms like Wikipedia. Wolniewicz-Slomka's study utilizes a historical analytical framework within the realm of sociology, focusing on the connection between collective memory and national identity. The author discovers that differences across three languages do exist, particularly in the language and the attention given to specific issues, indicating that Wikipedia editors are influenced by their cultural and social backgrounds. Compared to Wolniewicz-Slomka's findings, our study delves further by investigating a range of topics and languages. Our study finds that differences across languages in Wikipedia articles tend to reflect the most dominant intellectual views, introducing a type of algorithmic bias that limits users' perspectives to their own culture. Furthermore, our study addresses the potential pitfalls and dangers associated with language bias embedded on platforms like Google, ChatGPT, YouTube, and Wikipedia. By shedding light on the issues involving language bias, we hope to raise awareness and encourage the development of more inclusive algorithms to overcome language bias.

## Appendix 2  Additional Terms

### A special case of "socialism" on Google

The fall of the Soviet Union had a profound impact on the global perception of socialism. Following the collapse of the Soviet Union, socialism was widely viewed as a threat to individual freedom and was often associated with authoritarian governments. However, in recent years, there has been a shift in attitudes toward socialism, particularly in the U.S.

This shift is partly due to the influence of Bernie Sanders, a democratic socialist, who ran for the U.S. presidential campaign in 2016. His campaign, which focused on issues such as income inequality, universal healthcare, free college education, and a living wage for all workers, brought attention to the idea of socialism and generated a renewed interest in it as a solution to the problems facing the country today.

The shift in public perception of the term "socialism" is reflected in Google's English search results. The top-ranked websites now mention Bernie Sanders and his political position of democratic socialism, with titles "What Americans Must Know About Socialism"[60] and "What is Democratic Socialism?"[61] This suggests that the term has become more mainstream and widely discussed in recent years. Google's top-ranked search results might have a more negative portrayal of this term seven years ago before Sanders' presidential campaign. In this case, the dominant cultural phenomenon in this term is heavily influenced by a political figure whose campaign team might have invested millions of funds in advertising and promoting his idea.

### Liberalism on Wikipedia

Wikipedia has the "Liberalism" page available in 133 languages, and each language presents more or less an ethnocentric perspective on this subject matter. The English article has a long section on Keynesian economics. One of the key components of the New Deal implemented by President Franklin D. Roosevelt in the 1930s during the Great Depression was Keynesian liberalism, which promoted the idea that government intervention was necessary to stabilize the economy and address social welfare issues. President Lyndon Johnson continued to promote Keynesian liberalism after the New Deal, improving social welfare through programs like the Great Society and raising taxes to prevent inflation. The impact of Keynesian economics on the English-speaking community was significant, as it helped the U.S. and U.K. to recover from the depression. At the time, Britain had three million unemployed, and America had 15 million.

In the French article on liberalism, the focus is not solely on the modern interpretation of the term but rather on its historical roots and development. It traces its origins from antiquity to the Renaissance period, with a focus on French history. It also highlights the works of French theorists such as Montesquieu and Tocqueville and explores how the concept of liberalism has evolved over time and across different regions. The Italian article has a lengthy section on "Liberalism and Christianity" because liberalism can be seen as a threat to the catholic church. Hebrew has a section discussing the Zionist movement in Israel. The German article is much shorter than the French, Italian, and Hebrew ones. Due to Germany's loss in WWII, its post-WWII state was a liberal state and was occupied by the Allied forces consisting of troops from the U.S., U.K., France, and the Soviet Union. This might have influenced Germany's perception and approach to liberalism.

### Marxism on Wikipedia

On "Marxism," the English article has a section on "libertarian" and a section on "Humanist," but both sections do not appear in the "Marxism" articles in other languages. Libertarian

Marxism is a political and economic philosophy that combines Marxist analysis with a strong commitment to individual liberty. It is heavily influenced by Anglo-American thought, which stresses individual rights and freedoms. On the other hand, humanist Marxism, which emphasizes the importance of human agency and individual dignity, is rooted in the work of scholars such as Michel Foucault and Jacques Derrida, who have had a profound influence on American academic institutions.

The Russian article includes a section on the split of the Soviet Union and a lengthy section on Russian Marxist schools. Interestingly, Russian article begins with a section on the "Communist manifesto" which does not exist in the Chinese one. Since both Russia and China follow Marxism and the Communist manifesto is a founding pillar of the Communist ideology, a potential rationale to explain the absence of the Communist manifesto in the Chinese article is that in Mao era China, people read Mao Zedong Thought instead of the Communist manifesto because Mao realized that his 'Marxism' was different from the original Marxism.[12]

The first section of the German article on "Marxism" is Conceptual History which also does not exist in articles in other languages. Conceptual history (Begriffsgeschichte) is a field of study that originated in Germany but is not widely acknowledged in other language institutions. This section explores the evolution of the concept of "Marxism" over time, tracing its roots to the ideas of Karl Marx and Friedrich Engels and its development through the writings of subsequent Marxist thinkers. In comparison, the French article connects the topic back to political history and references French political contexts and theorists like Tocqueville, and Pierre Bourdieu. Though randomness is involved, variations of topic coverage in the German and French articles partially reflect the different intellectual focuses in studying and understanding Marxist ideology.

In general, we find that although randomness is involved in topic coverage, Wikipedia articles tend to reflect the dominant intellectual traditions and camps in different language communities. The dominant intellectual traditions and camps can relate to many factors, with the most prominent being academic scholars and institutions. But key political figures like Mao Zedong, John Maynard Keynes, and major social events like the Zion movement can also influence the breadth of topics in Wikipedia articles.

## Liberalism on YouTube

On "Liberalism," we find that French videos frequently connect liberalism with neoliberalism[62], but English videos do not mention neoliberalism at all (like English's Google search results). The Chinese and Korean videos are primarily talk shows and TV shows on liberalism, with few education-oriented channels. Japanese videos on '-ism' terms are dominated by one Japanese channel けんた国際政治チャンネル[63] (Kenta International Politics Channel), who has 1.62K subscribers. It is interesting to see that Russian videos associate liberalism with democracy and discuss whether it led to the dissolution of the Soviet Union[64], whereas English videos, mainly produced by big channels like CrashCourse and PragerU, focus on ideas like 'liberals vs. conservatives' and 'left vs. liberal.' A potential interpretation of the discrepancy here is that Russians tend to look at liberalism as a political system together with democracy in a broad scope, whereas Americans tend to be confused by the ideas of liberals, conservatives, and leftists. The fact that Russian videos focus on whether liberalism led to the collapse of the Soviet Union can be seen as a response to economic turmoil following the introduction of neoliberal economic policies advocated by American economists.[13]

## Cultural Revolution on YouTube

Likewise, "Cultural Revolution" videos vary greatly. In contrast to English videos that uniformly interpret this term as China's planetarian cultural revolution, Russian videos feature the early Soviet Union's cultural revolution, which was a period of radical social change that sought to break out of traditional feudal values to build a new modern state. The video discussing China's revolution is titled "Cultural Revolution in China." Likewise, French search results have a video about the Nazi cultural revolution. On the other hand, Asian language videos tend to be more negative than videos in European languages, presumably because they might feel this movement was closer to their life. Korean videos are titled "history of mass madness" and "worst agitation in history" rather than 'history of cultural revolution.' Vietnamese videos have titles like "Massacre in China," "Disaster in China," and "Big black spot in Chinese history." Compared to textual information on Google and Wikipedia, watching the top returned YouTube videos on the cultural revolution, viewers can perceive the ardent zeal and enthusiasm of the people during that era, and feel the tremendous power that ideology can wield over individuals.

---

[12] Mao Zedong's version of Marxism was based on the idea of a peasant rebellion, rather than on the traditional Marxist theories. According to Marxist theory, communism can only be established in countries that have reached a certain level of capitalism. However, at the time Mao came to power, China had not yet reached this level of capitalism. As a result, Mao's approach to implementing communism in China was based on the idea of a peasant rebellion, rather than the traditional Marxist model. This difference in ideology is one reason why Mao Zedong's ideas are often referred to as "Mao Zedong Thought," rather than simply "Marxism."

[13] Following the Soviet Union's dissolution in 1991, Russia transitioned from a state-controlled to a free-market economy – a process known as "shock therapy." This transition, implemented with price and trade liberalization, privatization, and the opening of markets to foreign investors, gave rise to a few oligarchs who amassed immense wealth and significant political influence, and contributed to an extreme economic inequality in Russia.

## Appendix 3  TF-IDF Metric

The formula of TF-IDF is the product between Term Frequency (TF) and Inverse Document Frequency (IDF).

TF, $tf(t, d)$, is defined as:

$$tf(t, d) = f_{t,d}, \qquad (1)$$

where $f_{t,d}$ is a number of occurrences of term $t$ in document $d$.

IDF, $idf(t, D)$, is defined as:

$$idf(t, D) = log\ log\ \frac{N}{count(d \in D : t \in d)}, \qquad (2)$$

where N is the total number of documents in the corpus, $count(d \in D : t \in d)$ is the total number of documents that contain the term $t$.

TF-IDF is the dot product between TF and IDF:

$$tfidf(t, d, D) = tf(t, d) \cdot idf(t, D) \qquad (3)$$

We can think of the IDF as a weighting factor that if a common word like "the" repeatedly appears in all documents, the IDF score for "the" would be ("$the$", $D$) $= log\ log\ \frac{1}{1} = 0$, which leads to 0 contribution in distinguishing a document from a corpus. For words that only appear in one document, it would have a high IDF score and weight as an important contributing factor in classifying a document. The TF-IDF metric is helpful in our study because it tells us the relative importance of a word in a particular language's content and can be used to compute content similarity.

The TfidfVectorizer function from the scikit-learn python library automatically converts each document in a corpus into a vector representation using TF-IDF. The vector representation for each document has a size of 1 x $V$, where $V$ is the vocabulary size of the entire corpus $D$, and each number in the vector is the TF-IDF score for term $t$.

Our study uses the TfidfVectorizer function to obtain document representation for each language content. For a Google search term like "Buddhism", corpus $D$ consists of translated website titles from $n$ languages, and each language content is considered as one document. The TfidfVectorizer function returns $n$ vector representations with respect to $n$ language documents, and we compute $n$ cosine similarity scores with respect to each language pair.

## Appendix 4 Buddhism-related Terms and Translations

| English | German | French | Italian | Spanish | Russian | Japanese | Korean | Chinese | Vietnamese | Thai | Nepali |
|---|---|---|---|---|---|---|---|---|---|---|---|
| Buddhism | Buddhismus | bouddhisme | buddismo | Budismo | буддизм | 仏教 | 불교 | 佛教 | đạo Phật | ศาสนาพุทธ | बुद्ध धर्म |
| Buddhist philosophy | Buddhistische Philosophie | Philosophie bouddhiste | Filosofia buddista | filosofía budista | Буддийская философия | 仏教哲学 | 불교철학 | 佛教哲学 | Triết học phật giáo | พุทธปรัชญา | बुद्ध दर्शन |
| Buddhist meditation | Buddhistische Meditation | Méditation bouddhique | Meditazione buddista | meditación budista | Буддийская медитация | 瞑想 | 명상 | 禅定 | Thiền định Phật giáo | การทำสมาธิแบบพุทธ | बौद्ध ध्यान |
| Gautama Buddha | Gautama Buddha | Bouddha Gautama | Gautama Buddha | Buda Gautama | Гаутама Будда | 釈迦牟尼仏 | 고타마 부처 | 释迦牟尼佛 | Đức phật | พระโคตมพุทธเจ้า | गौतम बुद्ध |
| Four Noble Truth | Vier edle Wahrheiten | Quatre nobles vérités | Quattro Nobili Verità | Cuatro nobles verdades | Четыре благородные истины | 四諦 | 네 가지 고귀한 진리 | 四谛 | Tứ diệu đế | อริยสัจ 4 | चतुरार्यसत्य |
| Karma in Buddhism | Karma-Buddhismus | bouddhisme karmique | Karma nel buddismo | karma en el budismo | Карма в буддизме | カルマ仏教 | 불교의 카르마 | 因果 佛教 | Nghiệp Phật | พุทธศาสนากรรม | कर्म बौद्ध |
| Reincarnation | Reinkarnation | Réincarnation | Reincarnazione | Reencarnación | Реинкарнация | 生まれ変わり | 재생 | 转世 | Đầu thai | การกลับชาติมาเกิด | पुनर्जन्म |
| Saṃsāra | Saṃsāra | Saṃsāra | Samsara | Saṃsāra | Сансара | 輪廻 | 윤회 | 輪迴 | Luân hồi | วัฏสงสาร | संसार |
| Lotus Sutra | Lotos-Sutra | Sūtra du Lotus | Sutra del Loto | Sutra del loto | Лотос Сутра | 妙法蓮華経 | 법화경 | 妙法莲华经 | Pháp hoa kinh | สัทธรรมปุณฑรีกสูตร | सद्धर्मपुण्डरीक सूत्र |
| Chögyam Trungpa | Chögyam Trungpa | Chögyam Trungpa | Chogyam Trungpa | Chogyam Trungpa | Чогьям Трунгпа | チョギャム・トゥルンバ | | 丘揚創巴 | Chögyam Trungpa | เชอเกียม ตรุงปะ | |
| Dalai Lama | Dalai Lama | Dalaï-lama | Dalai Lama | Dalai Lama | Далай Лама | ダライ・ラマ | 달라이 라마 | 达赖喇嘛 | Đạt-lai Lạt-ma | ดาไลลามะ | दलाई लामा |
| Kelsang Gyatso | Kelsang Gyatso | Kelsang Gyatso | Kelsang Gyatso | Kelsang Gyatso | Келсанг Гьяцо | ダライ・ラマ 7 世 | 켈상 가쵸 | 格桑嘉措 | Kelsang Gyatso | | |
| Ogyen Trinley Dorje | Orgyen Thrinle Dorje | Ogyen Trinley Dorjé | Ogyen Trinley Dorje | Ogyen Trinley Dorje | Огьен Тринле Дордже | 17 番目のカルマパ | 오겐 트린리 도르제 | 第十七世噶玛巴·伍金赤列多吉 | Karmapa thứ 17 | ออเกียน ทรินเลย์ ดอร์จี,องค์กรรมาปะที่ ๑๗ | |
| Thích Nhất Hạnh | Thích Nhất Hạnh | Thích Nhất Hạnh | Thích Nhất Hạnh | Thích Nhất Hạnh | Тич Нхат Хон | ティク・ナット・ハン | 틱낫한 | 释一行 | Thích Nhất Hạnh | ติช นัท ฮันห์ | |